# Dynamics of the nuclear gas and dust disc in the E4 radio galaxy NGC 7052

Frank C. van den Bosch[1] and Roeland P. van der Marel[1,2,3]

[1] *Sterrewacht Leiden, Postbus 9513, 2300 RA Leiden, The Netherlands (vdbosch@strw.strw.leidenuniv.nl)*
[2] *Institute for Advanced Study, Olden Lane, Princeton, NJ 08540, USA (marel@guinness.ias.edu)*
[3] *Hubble Fellow*

**ABSTRACT**
We present high spatial resolution ground-based broad-band imaging, H$\alpha$+[NII] narrow-band imaging and long-slit spectroscopy for the E4 radio galaxy NGC 7052, which has a nuclear dust disc. We detect ionized gas with a LINER spectrum, residing also in a nuclear disc. The gas rotates rapidly ($v_{\rm rot} \approx 250 \, {\rm km \, s^{-1}}$ at $1.5''$). The emission line widths increase towards the nucleus (nuclear FWHM $\approx 500-600 \, {\rm km \, s^{-1}}$).

The images are well fit by an axisymmetric model with the gas and dust in a disc of $\sim 1.5''$ radius ($\sim 340$ pc) in the equatorial plane of the stellar body, viewed at inclination $i \approx 70°$. The flux distribution of the ionized gas is very centrally concentrated, and is unresolved inside the seeing radius. The central cusp-steepness $\alpha$ of the stellar luminosity density ($j \propto r^\alpha$ for $r \to 0$) is not well constrained by the observed surface brightness distribution, because of the dust absorption.

We model the gas kinematics with the gas on circular orbits in the equatorial plane, with a local velocity dispersion due to turbulence (or otherwise non-gravitational motion). The circular velocity is calculated from the combined gravitational potential of the stars and a possible nuclear black hole. The observed gas rotation curve is well fit by a model with $\alpha = -1.3$, either with or without a black hole. Turbulent velocities $\gtrsim 300 \, {\rm km \, s^{-1}}$ must be present at radii $\lesssim 0.5''$ to fit the observed nuclear line widths. Seeing convolution of a Keplerian rotation curve around a $10^9 \, {\rm M}_\odot$ black hole can fit the observed widths without turbulence, but such a model predicts nuclear emission line shapes with pronounced peaks at $v \approx \pm 300 \, {\rm km \, s^{-1}}$, which are not observed. Models with both a black hole and gas turbulence can fit the data well, but the black hole is not required by the data, and if present, its mass must be $\lesssim 5 \times 10^8 \, {\rm M}_\odot$.

Although this upper limit is not very stringent, it is already $\sim 5$ times smaller than the black hole mass inferred for M87 from HST data. We show that HST observations of NGC 7052 should improve significantly the constraints on the mass of any possible black hole. Kinematic observations of nuclear gas discs are likely to become a widely used tool in the search for massive black holes in galactic nuclei. The modelling and analysis techniques presented here will be useful for the interpretation of such data.

## 1 INTRODUCTION

It is generally believed that active galaxies are powered by the presence of a massive central black hole (e.g., Rees 1984). In fact, black holes are expected to be present in the centres of many quiescent galaxies as well (e.g., Chokshi & Turner 1992). However, direct dynamical evidence for the presence of black holes in individual galaxies is scarce. Stellar kinematic studies (e.g., Kormendy 1993; van der Marel et al. 1994a,b,c) have provided tentative evidence for black holes in a handful of nearby galaxies. However, the case is clear-cut for none of them, due to finite spatial resolution and lack of knowledge on the shape of the stellar orbits. An alternative is to study the dynamics of ionized gas, if present. This can be very complicated if the gas morphology is irregular or unresolved, or if there are significant contributions from, e.g., inflow or outflow. This is often the case for the narrow-line and broad-line regions of active galactic nuclei. However, if there is observational evidence that the gas moves on (approximately) circular orbits in a disc, the analysis is straightforward and less complicated than for the stellar kinematics. M87 provides a good example. Van der Marel (1994c) detected high velocities near the nucleus which he modelled as gravitational motion around a black hole. Subsequent observations with the Hubble Space Telescope (HST) by Ford et al. (1994) and Harms et al. (1994) resolved the gas spatially, and showed it to be in a rapidly rotating disc. The observed gas velocities in M87 provide the currently best evidence for a black hole in any galactic nucleus (with the possible exception of our own galaxy).

NGC 7052 is an E4 galaxy with apparent magnitude $m_B = 13.4$ (de Vaucouleurs et al. 1991), at heliocentric velocity $v_{\rm hel} = 4687 \, {\rm km \, s^{-1}}$ (Faber et al. 1989). It is a known radio source with a core and jet, but no lobes (Morganti et al. 1987). The radio axis is offset from the photometric major axis of the galaxy by $\sim 50°$ (Möllenhoff, Hummel & Bender 1992). It is also an infrared source. It was detected in all four IRAS bands and has a far-infrared luminosity of $\sim 10^{10} \, L_\odot$ (Impey, Wynn-Williams & Becklin 1990). Optical images show absorption by dust near the nucleus. Very high spatial resolution (seeing FWHM $0.33''$) CFHT images of the nucleus were presented by Nieto et al. (1990). These show that the dust resides in a lane aligned with the major axis (see also Figure 1), suggesting it to be in a nearly edge-on disc with $\sim 1.6''$ radius. Molecular gas was detected in NGC 7052 using the CO (1-0) line (Wang et al. 1992). Atomic gas has not been detected ($M_{\rm HI} < 5 \times 10^8 \, M_\odot$; Dressel, Bania & O'Connell 1982). Morganti, Ulrich & Tadhunter (1992) observed NGC 7052 through a narrow-band filter centred on H$\alpha$ and the adjacent [NII] lines. Ionized gas was detected around the nucleus, but its morphology was unresolved at the $\sim 2''$ resolution of their observations.

The presence of a nuclear dust disc in NGC 7052 suggests that the ionized gas might also reside in a disc, since the occurrence of dust and gas in elliptical galaxies are often closely coupled (Bertola 1992; Buson et al. 1993; Goudfrooij et al. 1994). If so, observations of the gas kinematics could yield useful constraints on the nuclear mass distribution. For this reason we observed NGC 7052 in August 1992 with the 4.2m William Herschel Telescope (WHT) on La Palma. Broad-band images (Section 2) and long-slit emission- and absorption-line spectra (Section 4) were obtained in excellent seeing conditions (FWHM $\approx 0.6''$). A narrow-band image was later also obtained (Section 3). Our observations show that the ionized gas indeed resides in a rotating disc. Interpretation of the observed motions requires careful modelling to correct for the effects of seeing and dust absorption. In Section 5 we model the stellar light distribution, the geometry of the ionized gas, and the geometry and optical depth of the dust. We thus fit the broad- and narrow-band images and the integrated emission line fluxes derived from the spectra. In Section 6 we model the observed gas kinematics. We show that a central black hole is not required to fit the data, but that some form of gas turbulence is. We put an upper limit on the mass of any possible black hole (or other dark compact nuclear mass). For our best-fitting models we calculate predictions for future spectroscopic observations with the HST. Conclusions are presented in Section 7.

Values quoted throughout this paper are based on a Hubble constant $H_0 = 100 \, h \, {\rm km \, s^{-1} \, Mpc^{-1}}$ with $h = 1.0$. The particular choice adopted for $h$ does not influence the predictions of any of our models. It merely sets the length, mass and luminosity scale of the models in physical units.

**Figure 1.** $R$-band image of the central region of NGC 7052, obtained with seeing FWHM $S = 0.67''$. The directions of North and East are as indicated. There is a dust lane in the inner $2''$, aligned with the major axis of the galaxy.

Specifically, distance, length scales and masses scale as $h^{-1}$, while mass-to-light ratios scale as $h$.

## 2 BROAD-BAND IMAGING

### 2.1 Data acquisition and reduction

Three $R$-band images of NGC 7052 were obtained with the GEC5 CCD at the Cassegrain focus Auxiliary Port of the WHT. The CCD pixels are $0.10'' \times 0.10''$, with a $38'' \times 58''$ field of view. The image reduction was done using standard IRAF tasks.

The seeing point spread functions (PSFs) of the observations were determined from bright stars on the images. The PSF shape could be well described by a sum of two Gaussians:

$${\rm PSF}(r) = A_1 e^{-r^2/2\sigma_1^2} + A_2 e^{-r^2/2\sigma_2^2}, \qquad (1)$$

with fixed ratios of the dispersions, $\sigma_2/\sigma_1 = 1.92$, and amplitudes $A_2/A_1 = 0.29$, of the two components. The PSF is normalized for $A_1 = 0.07743/\sigma_1^2$, and is fully specified by its FWHM $S = 2.644 \, \sigma_1$.

Our best image is a 600 sec. exposure with seeing FWHM $S = 0.67'' \pm 0.02''$. The analysis below is based on this image. The other two images were not used because of poor signal-to-noise ratio $(S/N)$ and poor seeing, respectively. Figure 1 shows the central region of NGC 7052. There is a sharp dust lane aligned with the major axis of the galaxy, in agreement with the (higher spatial resolution) image of Nieto et al. (1990).

The WHT autoguider, which was locked onto a bright star during each exposure, provides an independent measure of the seeing FWHM, $S_{\rm auto}$. It does not properly take the full PSF shape (1) into account though, and its estimates must be calibrated. Our images imply $S/S_{\rm auto} \approx 0.76$, in good agreement with the calibration derived by van der Marel et al. (1994a). Calibrated autoguider estimates are used in Section 4 to estimate the seeing FWHM for our WHT spectra.

**Figure 2.** $R$-band surface brightness, ellipticity, major axis position angle and $\cos 4\theta$ Fourier coefficients of the isophote deviations from ellipses, determined from the image in Figure 1. The $\cos 3\theta$, $\sin 3\theta$ and $\sin 4\theta$ coefficients were found to be close to zero, and are not shown. The abscissa is the radius along the major axis, measured with respect to the centre of the outer isophotes. The observed surface brightness at radii $< 2.5''$ cannot be meaningfully approximated by ellipses. The stellar surface brightness and dust geometry in this nuclear region are modelled in Section 5.

### 2.2 Isophotal Analysis

We analysed the isophote shapes with the ellipse fitting software of Jedrzejewski (1987), in the manner described by van den Bosch et al. (1994). Figure 2 shows the $R$-band surface brightness, ellipticity ($\epsilon$), major axis position angle (PA) and $\cos 4\theta$ Fourier coefficients of the isophote deviations from ellipses, as functions of radius along the major axis.

Accurate isophotal parameters could not be determined at radii $\gtrsim 10''$, because the galaxy was offset towards one corner of the CCD. This allows an accurate sky subtraction (using the measured brightness in the opposite corner of the CCD) at the cost of a more limited radial extent of the data. No isophotal results are shown for radii $\lesssim 2.5''$, where the isophotes cannot be meaningfully approximated by ellipses. Models for the stellar surface brightness at small radii, and for the geometry and optical depth of the dust are discussed in Section 5.

Between $r = 2.5''$ and $r = 10''$ the ellipticity of the isophotes increases from $\sim 0.3$ to $\sim 0.4$. The position angle is constant and the higher-order Fourier terms are close to zero. These results are consistent with previous studies by Bender, Döbereiner & Möllenhoff (1988) and Nieto et al. (1991). Both find that the ellipticity increases further to $\epsilon \approx 0.55$ at $r = 50''$. There is no agreement in the literature on the isophote deviations from ellipses for $r \gtrsim 20''$. Bender et al. find the isophotes to be boxy (negative $\cos 4\theta$ coefficient), Nieto et al. find the isophote deviations to be close to zero, while Ebneter, Djorgovski & Davis (1988) argue that NGC 7052 might be an S0 (which would seem to be consistent with the increasing ellipticity with radius; see the arguments in van den Bosch et al. 1994).

### 3 NARROW-BAND IMAGING

Marcella Carollo kindly obtained narrow-band images of NGC 7052 for us in July 1994, using the 1.5m Danish Telescope at ESO, La Silla. The CCD pixels are $0.377'' \times 0.377''$, with a $6.4' \times 6.4'$ field of view. Two 30 min. exposures were taken, one 'on-band' and one 'off-band'. The on-band filter covered the region around H$\alpha$ ($\lambda_{\rm rest} = 6562.68$Å) and the adjacent [NII] lines ($\lambda_{\rm rest} = 6548.03, 6583.41$Å), at the redshift ($\Delta\lambda \approx 102$Å) of NGC 7052. The off-band filter was centred on $\lambda \approx 6550$Å. Both filters have an effective width of $\sim 60$Å.

After bias, dark current, cosmic ray and sky subtraction, and flat-fielding, the frames were shifted to a common coordinate system using stars as reference points. No rotation of the frames was necessary. After measuring the seeing FWHM from stellar images, the best seeing frame was convolved with a Gaussian to yield an 'effective' seeing FWHM equal to the seeing FWHM for the worst seeing frame ($S = 1.41''$). Subsequently, the images were scaled to the same surface brightness at radii $\gtrsim 10''$ (were there is no line emission according to our spectra in Section 4.1), and the off-band image was subtracted from the on-band image.

The resulting H$\alpha$+[NII] image is shown in Figure 3. The innermost gas contours are flatter than the stellar contours, in spite of the relatively poor seeing of the narrow-band image (which tends to make the gas contours rounder). Hence, the ionized gas most likely resides in a disc. The major axes of the stellar and gas components are aligned within $2°$. The centre of the H$\alpha$ + [NII] emission is offset from the centre of the outer isophotes of the galaxy by $\sim 0.4''$. It is shown below (Section 5) that this can well explained as due to absorption of part of the line emission by the dust, which itself is not entirely symmetrically distributed with respect to the centre. There might be an additional contribution from the fact that the ionized gas rotates rapidly (Section 4.2), causing the transmission of the emission line flux by the on-band filter to vary with position.

**Figure 3.** Map of the Hα+[NII] emission (thick contours) overplotted on the observed broad-band surface brightness distribution from Figure 1 (thin contours). The narrow-band image was obtained in relatively poor seeing (FWHM $S = 1.41''$). The innermost gas contours are none the less flatter than the contours of the stellar component, indicating that the gas most likely resides in a disc. The offset between the centre of the Hα+[NII] emission and the centre of the outer stellar isophotes can be well understood as due to absorption of part of the line emission by a slightly asymmetric dust distribution (cf. Figures 8 and 10 below).

## 4 LONG-SLIT SPECTRA

Long-slit spectra of NGC 7052 were obtained with the R1200R grating and the EEV3 CCD on the red arm of the ISIS spectrograph of the WHT. The dispersion is 0.38 Å/pixel, the spatial scale $0.33''$/pixel. A slit width of $0.8''$ was used, yielding an instrumental velocity dispersion $\sigma_{\rm instr} \approx 10 \,{\rm km\,s^{-1}}$. Exposures of arc lamps were taken before and after each galaxy exposure to allow accurate wavelength calibration. The galaxy was centred on the slit at the beginning of each exposure using a TV camera with an $R$-band filter. The autoguider was equipped with an $I$-band filter. Differential atmospheric refraction between the $R$ and $I$-band is negligible ($\lesssim 0.1''$). The data were reduced as in van der Marel et al. (1994a). Standard steps included: bias subtraction, flat-fielding, cosmic ray removal, wavelength calibration, logarithmic wavelength rebinning, rebinning to correct for S-distortion, sky subtraction and dark current subtraction. The wavelength dependence of the efficiency of the telescope-detector combination was corrected using observations of a spectrophotometric standard star.

### 4.1 Absorption-line spectra: stellar kinematics

One 50 min. major axis (PA = 65°) absorption-line spectrum was obtained from 8390Å to 8820Å, covering the Ca II IR triplet region. The seeing FWHM was $S = 0.84''$ (obtained from the autoguider estimate; see Section 2). To obtain sufficient $S/N$ for a stellar kinematic analysis, all spectra along the slit with $-5'' \leq r \leq 5''$ were summed. This spectrum was then analysed with the pixel space fitting method of van der Marel (1994c), using the spectrum of a K5 III star obtained with the same instrumental setup as template. This yields an 'average' stellar velocity dispersion for the central region of NGC 7052: $\sigma = 266 \pm 26 \,{\rm km\,s^{-1}}$. This is consistent with the higher quality data of Wagner, Bender & Möllenhoff (1988), who inferred a central velocity dispersion of $\sim 270 \,{\rm km\,s^{-1}}$. They found that the stellar rotation velocities are comparable along the major and minor axes, indicative of triaxiality, but that the overall contribution of rotation to the dynamical support of the system is negligible ($|V_{\rm rot}/\sigma| \lesssim 0.2$). Unfortunately, the $S/N$ of our absorption-line spectrum is too low to study the variation of the stellar kinematics as a function of radius.

### 4.2 Emission-line spectra: ionized gas kinematics

One 30 min. major axis (PA = 65°) and one 30 min. minor axis (PA = 155°) spectrum were obtained from 6435Å to 6880Å, around Hα+[NII]. The seeing FWHM was $S = 0.57''$ for the major axis spectrum and $S = 0.84''$ for the minor axis spectrum (obtained from the autoguider estimates). Hence, the major axis spectrum was taken under very good seeing conditions, although not exceptional for La Palma (see, e.g., Zeilinger, Møller & Stiavelli 1993).

We performed the following test to verify the seeing estimate for our major axis spectrum. The $R$-band image in Figure 1 was 'overlaid' with a hypothetical slit along the major axis, with the same width and pixel size as for our spectroscopic observations. Upon summing the $R$-band intensity over each pixel, this yields the predicted intensity profile along the slit for spectroscopic observations with $S = 0.67''$ (the seeing FWHM for the $R$-band image). This predicted profile was found to be less centrally peaked than the observed intensity profile along the slit, consistent with the fact that the major axis spectrum had $S < 0.67''$.

To study the emission lines one needs to subtract the spectrum of the underlying stellar component. Fitting this component merely with a low-order polynomial can introduce systematic errors (e.g., Goudfrooij et al. 1994). Hence, we model the stellar component as the convolution of a template spectrum (the spectrum of a K3 III star obtained with the same instrumental setup) with a Gaussian velocity distribution. The latter was taken to be the same as determined in Section 4.1. This is adequate, because the stellar velocity dispersion does not vary much, and the stellar rotation velocities are small, over the spatial region where the emission lines are observed (Wagner et al. 1988).

No emission lines were detected outside the central $2''$, in agreement with Figure 3. Figure 4 shows the emission-line spectra for three representative radii along the major axis. There are no residual absorption lines, indicating that the starlight has been properly subtracted. In addition to Hα and the [NII] lines, we detect the [SII] lines with $\lambda_{\rm rest} = 6717.47, 6730.85$Å. The nuclear line ratios, $[{\rm NII}]_{6583}/{\rm H}\alpha \approx 1.7$ and $[{\rm SII}]_{6717+6731}/{\rm H}\alpha \approx 1.0$ are characteristic for a low ionization emission-line region (LINER), and are significantly larger than those seen in HII region-like objects (Veilleux & Osterbrock 1987).

The peak [NII] 6583Å line flux in our nuclear spectrum is $1.8 \times 10^{-16} \,{\rm erg\,cm^{-2}\,s^{-1}\,Å^{-1}}$. The integrated Hα+[NII] flux summed over all radii along the major axis is $1.8 \times 10^{-14} \, \mathrm{erg \, cm^{-2} \, s^{-1}}$. Morganti et al. (1992) measured a total H$\alpha$+[NII] flux for NGC 7052 of $2.4 \times 10^{-14} \, \mathrm{erg \, cm^{-2} \, s^{-1}}$.

**Figure 4.** Emission line spectra at three representative radii along the major axis (0.61″, −0.05″, −0.71″), after subtraction of the stellar contribution to the spectrum. The spectra are characteristic for a LINER. The vertical scale is in arbitrary units.

To determine the ionized gas kinematics we study the H$\alpha$ and [NII] emission lines. The [SII] lines are not used, because of their lower $S/N$. We obtain smooth fits to the H$\alpha$ and [NII] lines using a two step process. In each step a suitably parametrized function is fit to the data using the Levenberg-Marquardt method (e.g., Press et al. 1992). The $\chi^2$ quantity that is minimized takes read-out and photon noise into account. In the first step all three lines are assumed to be Gaussian, and the three Gaussians are fit to the data simultaneously. The ratio of the amplitudes of the [NII] lines is fixed to 3, since this is the ratio of their transition probabilities (Osterbrock 1989). Since both [NII] lines originate from the same gas component, their velocity dispersions are constrained to be equal, and their velocity difference is fixed to $\Delta v = c\Delta \ln\lambda = 1615.5 \, \mathrm{km \, s^{-1}}$. In the second step each line is fit separately using a function that is more general than a single Gaussian, namely the sum of two Gaussians. This has no physical significance, but is merely a convenient fitting function. Before fitting either line the other two lines are subtracted, using the Gaussian fits obtained in the first step. This procedure works well since there is only little overlap between different lines (Figure 4). Figure 5 shows the best double Gaussian fits to the major axis H$\alpha$ lines after subtraction of Gaussian fits to the [NII] lines, to illustrate the procedure.

From the best-fitting double Gaussian to each line we determined the velocity moments up to order 2 (the integrated line flux, mean velocity and velocity dispersion) and their formal errors. These are convenient quantities for comparison to the predictions of theoretical models. (Alternatively, the width of the emission lines could have been characterized by their FWHM, rather than by their dispersion $\sigma$. For a Gaussian profile these are related by FWHM $= \sqrt{8 \ln 2} \, \sigma$.) Figure 6 shows the results for the major and minor axes. The two individual [NII] lines yielded consistent results. The results shown in Figure 6 are for the line at $\lambda_{\mathrm{rest}} = 6583.41$Å, which has the highest $S/N$. The ionized gas rotates rapidly along the major axis, reaching $\sim 250 \, \mathrm{km \, s^{-1}}$ at $r = 1.5''$. The velocity dispersion of the ionized gas has a central peak. There are some differences between the results for H$\alpha$ and [NII]. The flux distribution of [NII] is more centrally peaked than that of H$\alpha$. Along the major axis, the rotation velocities of [NII] are lower and the velocity dispersions higher than those of H$\alpha$.

**Figure 5.** Observed H$\alpha$ emission lines at different radii along the major axis (thin curves), after subtraction of Gaussian fits to the neighbouring [NII] lines. Thick curves are the best-fitting double Gaussians. The shift of the lines with radius indicates rapid rotation of the gas. The velocity dispersion of the gas peaks in the centre.

**Figure 6.** Integrated flux (in arbitrary units), rotation velocity and velocity dispersion of the ionized gas along the major and minor axes of NGC 7052. Results for H$\alpha$ and [NII] are indicated with different symbols. Error bars are shown for H$\alpha$ only, to avoid confusion. The errors on the results for [NII] are of similar magnitude.

## 5 MODELLING THE STELLAR LIGHT DISTRIBUTION AND GAS AND DUST GEOMETRY

### 5.1 Model assumptions

We assume for the three-dimensional stellar luminosity density $j$: (i) that it is oblate axisymmetric; (ii) that the isoluminosity spheroids have constant flattening $Q$ as function of radius; and (iii) that $j$ can be parametrized as:

$$j(R,z) = j_0(m/b)^\alpha [1 + (m/b)^2]^\beta, \quad m^2 \equiv R^2 + z^2 Q^{-2}, \quad (2)$$

where $(R, z)$ are the usual cylindrical coordinates, and $j_0$, $b$, $\alpha$ and $\beta$ are free parameters. When viewed at inclination angle $i$ in the absence of dust absorption, the projected intensity contours are ellipses with axial ratio $Q'$, with

$$Q'^2 \equiv \cos^2 i + Q^2 \sin^2 i. \quad (3)$$

Models with the same $Q'$ but different $i$ have the same projected intensity apart from a normalization constant. The projected intensity must generally be calculated numerically. Simple analytical expressions exist only for special combinations of the model parameters.

The assumption of axisymmetry for the luminosity density cannot be entirely correct, given that stellar streaming is observed along both the major and minor axes (Wagner et al. 1988). However, there are several indications that NGC 7052 might in fact be close to axisymmetric, or at least only mildly triaxial. There is no isophotal twisting (Figure 2), the dust (Figure 1) and ionized gas (Figure 3) are aligned with the major axis of the galaxy, and there is no significant rotation of the gas along the minor axis (Figure 6).

The assumption of constant flattening for the luminosity density was adopted because it simplifies the dynamical modelling. However, it does not fit the observations very well. The observed ellipticity of the isophotes increases steadily with radius beyond $\sim 2.5''$ (Figure 2). None the less, we consider our assumption to be adequate. First, the main thing we want to include in the model is that NGC 7052 is flattened; changes in flattening are only a second order effect. Second, what is most important for our dynamical modelling is the flattening in the central $\sim 2.5''$, which is poorly constrained observationally because of the absorption by dust. Adopting models with variable flattening would not resolve this degeneracy. Third, the influence of ellipticity changes can be assessed by comparing the dynamical predictions of models with different constant flattening. We did this, and found no significant change in any of the major conclusions of our paper. In the following we adopt $Q' = 0.7$

**Figure 7.** The top panels show the observed minor axis $R$-band surface brightness profile in the direction least affected by dust absorption, as function of distance to the centre of the outer isophotes. The solid curves are fits of the seeing convolved projected intensity of our models to the data with $r > 1''$. The dashed curves show the predictions before seeing convolution. Two representative choices for the cusp steepness $\alpha$ are shown, $\alpha = 0$ (a constant surface brightness core) and $\alpha = -1.3$ (a central power law cusp). The lower panels show the residuals between the model and the observations, as resulting from dust absorption. The cusped model requires a larger optical depth for the dust, and a dust distribution extending to a larger radius.

(i.e., $\epsilon = 0.3$), which fits the observed ellipticity at $r = 2.5''$.

The parametrization (2) for the luminosity density has no particular physical significance. It was chosen because it reproduces the generic behaviour of the surface brightness profiles of elliptical galaxies observed with the HST: a central power law cusp in the inner parts with a smooth change at some intermediate radius to a steeper power law fall-off at larger radii (e.g., Crane et al. 1993; Ferrarese et al. 1994; Kormendy et al. 1994). Models with $\alpha = 0$ have a constant luminosity core. Models with $\alpha > -1$ have a constant surface brightness core when projected onto the sky. Models with $\alpha < -1$ have a central power-law surface brightness cusp when projected onto the sky.

For lack of information on the distributions of the ionized gas and dust along the line of sight, we assume both components to be infinitesimally thin, residing in the equatorial plane of the galaxy. Furthermore, we assume the ionized gas flux $F$ to be circularly symmetric, $F = F(R)$. We consider distributions of the form:

$$F(R) = \begin{cases} I_1 \exp(-R/R_1) + I_2 \exp(-R/R_2), & (R \leq \bar{R}) \\ 0, & (R > \bar{R}) \end{cases} \quad (4)$$

This parametrization was chosen because with a proper choice for the free parameters $I_1$, $I_2$, $R_1$, $R_2$ and $\bar{R}$, it can adequately fit our data. It has no further significance. The flux distributions of H$\alpha$ and the [NII] lines are different (Figure 6), and we thus use different sets of parameters to describe them. The optical depth $\tau$ of the dust is not constrained to be circularly symmetric, because it is not a priori clear whether the dust resides in a disc (or ring), or in a single filament. We determine the dust distribution from the data using an unparametrized technique (Section 5.3).

The observed emission line flux $F_{\rm obs}$ need not represent the intrinsic flux $F$, since part of it might have been absorbed by dust. The fraction of the emission line flux that is absorbed depends on the distributions of the gas and dust in the direction perpendicular to the equatorial plane, which we do not model. We therefore include absorption of line emission by dust phenomenologically, by considering two limiting cases: $F_{\rm obs} = F$ ('gas in front of the dust') and $F_{\rm obs} = e^{-\tau} F$ ('gas behind the dust'). In the latter case, $\tau$ is assumed to be the same as in the $R$-band. The difference in extinction between the wavelengths of the $R$-band and H$\alpha$+[NII] is negligible (Osterbrock 1989).

Our model for the gas and dust in NGC 7052 is not elaborate. However, it captures the essential physics of the problem, and as will be shown below, it allows good fits to be obtained to the observed dust and gas morphology, including its asymmetries.

### 5.2 The minor axis surface brightness profile

A useful first constraint on the parameters of the luminosity density $j$ is obtained by fitting to the observed minor axis surface brightness profile in the direction least affected by dust absorption ('up' in Figure 1). Figure 7 shows this profile as function of distance to the centre of the outer isophotes. This centre is not the point of maximum surface

**Figure 8.** Thick contours show the (seeing-deconvolved) dust optical depth, obtained with the algorithm in Section 5.3, for $i = 70°$ and two representative values of the cusp-steepness $\alpha$. The left panel shows the contours of the optical depth for $\alpha = 0$ (contours at 0.2, 0.5, 0.8 and 1.1), while the right panel shows the more symmetrical optical depth distribution for $\alpha = -1.3$ (contours at 0.2, 0.5, 0.9, 1.5 and 3.0). Thin contours show the observed broad-band surface brightness distribution.

brightness, due to dust absorption. The cusp-steepness $\alpha$ is thus not well constrained.

The profile outside $\sim 1''$ shows little sign of being influenced by dust absorption. For any fixed choice for $\alpha$, the parameters $j_0$, $b$ and $\beta$ can thus be determined by fitting the seeing convolved projected intensity of the model to the data outside $\sim 1''$. Figure 7 shows the best fits for two representative cases: $\alpha = 0$ and $\alpha = -1.3$. The residuals of the fit in the central arcsec are largest for the model with $\alpha = -1.3$. This model has more light near the centre, and thus requires a larger optical depth for the dust to fit the data (see Figure 9 below).

The shapes of the fits in Figure 7 do not depend on the assumed inclination angle $i$. There is thus a two-parameter family of plausible models for the luminosity density, with $\alpha$ and $i$ as free parameters.

### 5.3 Determining the dust distribution

Given a model for the luminosity density $j$, the optical depth of the dust can be determined with the following algorithm.

Let the $w$-axis be along the line-of-sight and let the $z$-axis (the symmetry axis of the galaxy) be oriented such that the observer sees the region $z > 0$ in front of the dust disc and the region $z < 0$ behind the dust disc. Define

$$I_+(x,y) = \int_{z>0} j \, dw, \qquad I_-(x,y) = \int_{z<0} j \, dw, \qquad (5)$$

where $(x, y)$ are Cartesian coordinates on the sky. The projected intensity is

$$I(x,y) = I_+(x,y) + I_-(x,y) e^{-\tau(x,y)}, \qquad (6)$$

where $\tau(x, y)$ is the optical depth of the dust as seen projected onto the sky. The observed intensity is obtained by convolving with the normalized seeing PSF, $P(r)$:

$$I_{\text{obs}}(x,y) = \int I(x',y') P(|\vec{r}\,' - \vec{r}|) \, dx' \, dy', \qquad (7)$$

where $\vec{r} = (x, y)$ and $\vec{r}\,' = (x', y')$.

For any given inclination and model for $j$ one can evaluate $I_+$ and $I_-$, whereas $I_{\text{obs}}$ and $P(r)$ are known from observations (Section 2). To recover the unknown $\tau(x, y)$ we define and evaluate

$$\tilde{I}(x,y) \equiv I_{obs}(x,y) - \int I_+(x',y') P(|\vec{r}\,' - \vec{r}|) \, dx' \, dy'. \qquad (8)$$

Combination of equations (6)–(8) yields

$$\tilde{I}(x,y) = \int \mathcal{T}(x,y) P(|\vec{r}\,' - \vec{r}|) \, dx' \, dy', \qquad (9)$$

where we defined

$$\mathcal{T}(x,y) \equiv e^{-\tau(x,y)} I_-(x,y). \qquad (10)$$

Recovering the unknown $\mathcal{T}(x, y)$, and hence $\tau(x, y)$, from equation (9) is a seeing deconvolution problem. To solve it, we implemented a version of the Richardson-Lucy algorithm (Lucy 1974). Convergence was usually achieved after $\sim 40$ iterations. Results are discussed in Section 5.5 (see Figure 8).

### 5.4 Modelling the ionized gas distribution

The observed ionized gas distribution on the plane of the sky determines the inclination angle $i$ of our models, because the intrinsic distribution is assumed to be circularly symmetric in the equatorial plane. Recovering $i$ from Figure 3 is not straightforward though, because: (i) the observed distribution has been convolved with a circular seeing PSF; (ii) part of the line-emission flux might have been absorbed by dust; (iii) the transmission of the emission line flux by the on-band filter used in the narrow-band imaging varies with position, due to the rotation of the gas; and (iv) the flux distributions for H$\alpha$ and the [NII] lines are different. To circumvent problems (iii) and (iv), we determine $i$ by fitting to the fluxes observed in our major and minor axis spectra (Figure 6), rather than to our narrow-band image (Figure 3). The following approach was adopted to take (i) and (ii) into account.

**Figure 9.** The dust optical depth at the projected position of the galaxy centre, $\tau_{\rm cen}$, as function of the cusp-steepness $\alpha$. Models with a steeper power-law cusp (smaller $\alpha$) imply a larger optical depth for the dust.

Consider a fixed choice for $\alpha$, and a trial value for $i$. The parameters $j_0$, $b$ and $\beta$ of the luminosity density are fitted as in Section 5.2. Subsequently, the optical depth $\tau(x, y)$ is determined as in Section 5.3. A function $F(R)$ of the form (4) is then fitted to the integrated line flux observed in the major axis spectrum, taking into account the seeing convolution, the slit width and the binning into pixels along the slit. Given the fitted form for $F(R)$, the predicted fluxes are calculated for the minor axis spectrum, and compared to the data. This procedure is repeated for different values of $i$. The one that yields the best fit to the minor axis flux distribution is adopted as final estimate.

### 5.5 Results

We calculated the best-fitting inclination for H$\alpha$ and [NII] separately, and for both assumptions for the absorption of the emission-line flux by dust, $F_{\rm obs} = F$ and $F_{\rm obs} = e^{-\tau}F$ (as discussed in Section 5.1). In addition, we examined different values of $\alpha$ (which comes in because it determines $\tau$). The best fit (see Figure 10 below) was without exception $i = 70° \pm 5°$.

Having fixed the inclination, $\tau(x, y)$ depends on $\alpha$ only. Figure 8 shows the result of calculating it with the algorithm of Section 5.3, for two representative cases: $\alpha = 0$ and $\alpha = -1.3$. For $\alpha = 0$ the optical depth is very asymmetrical, the point of maximum $\tau$ being offset from the centre of the outer isophotes by $\sim 0.7''$ along the major axis, and $\sim 0.3''$ along the minor axis. The optical depth distribution for $\alpha = -1.3$ is more symmetric (though not entirely), and the $\tau$-contours are more nearly elliptical. Assuming the dust distribution in the equatorial plane to be circularly symmetric, the outermost $\tau$ isophote indicates $i \approx 72°$, close to the value derived from the ionized gas distribution. This was not imposed a priori, and hence provides a successful consistency check on the model.

Figure 9 shows the required optical depth in the centre, $\tau_{\rm cen}$, as function of the assumed cusp steepness $\alpha$. Models with smaller $\alpha$ have more light near the centre, and thus require a larger optical depth for the dust to fit the data. Unfortunately, we only have broad-band images in one band. Independent constraints on the optical depth of the dust (and hence on $\alpha$) from the observed reddening are thus not available. The observed infrared emission from NGC 7052 (part of which might in fact be non-thermal) can be used to estimate the total mass of dust in NGC 7052 (Jura 1986), but this does not yield useful constraints on $\alpha$.

HST observations have revealed that essentially all elliptical galaxies have central power-law cusps, rather than constant surface brightness cores (Crane et al. 1993; Ferrarese et al. 1994). This favours models with $\alpha < -1$ as most plausible explanation of our data. Such models also imply a more symmetrical dust distribution than models with, e.g., $\alpha = 0$. In the following we restrict ourselves to the model with $\alpha = -1.3$ (which, in the absence of dust absorption, corresponds to $I \propto r^{-0.3}$ in projection). As will be shown (see Figure 12 below), this yields the best fit to the observed major axis gas rotation curve. In addition, it is a more-or-less typical value for bright elliptical galaxies (Kormendy et al. 1994). For example, M87 has $\alpha = -1.26$ (Lauer et al. 1992).

With $i = 70°$ and $\alpha = -1.3$ the best fitting values for $j_0$, $b$ and $\beta$ are: $j_0 = 6.0\,{\rm L}_\odot\,{\rm pc}^{-3}$, $b = 2.35''$ and $\beta = -0.58$. We henceforth refer to this model as our 'standard model'. Its total luminosity is formally infinite. This is no great drawback. The agreement between the model and the data only breaks down at large radii, where no kinematic data are available anyway.

Figure 10 shows the best fits to the integrated H$\alpha$ fluxes along the major and minor axes for our standard model, with $\tau(x, y)$ as in Figure 8, and $F(r)$ as in equation (4). Models that assume $F_{\rm obs} = F$ and $F_{\rm obs} = e^{-\tau}F$ are both shown. The latter assumption provides the better fit. It reasonably well explains the slight asymmetries in the flux distribution, as due to absorption by the somewhat asymmetric dust distribution. In the following we restrict ourselves to this case. However, adopting the alternative assumption, $F_{\rm obs} = F$, would not change the main conclusions of our paper.

The ($F_{\rm obs} = e^{-\tau}F$) model for the H$\alpha$ fluxes in Figure 10 has $R_1 = 0.1''$, $R_2 = 1.0''$, $\bar{R} = 1.5''$ and $I_1/I_2 = 120$. The ratio of the total flux in the first exponential component of $F(R)$, $F_1$, to that in the second component, $F_2$, is $F_1/F_2 = 2.7$. Hence one can think of the flux as being the sum of a spatially unresolved component ($R_1 \ll$ seeing FWHM) which contains most of the flux, and a resolved extended component. The observed flux in the [NII] lines is more centrally peaked than that in H$\alpha$ (Figure 6). It can be fit with a similar flux distribution as above, but with $F_1/F_2 = 13.5$. The spatially unresolved component might be a so-called 'narrow-line region'. Such regions are seen in many active galactic nuclei, and have a typical size of $\sim 100$ pc (e.g., Osterbrock 1991). This would indeed be unresolved at the distance of NGC 7052 (100 pc = $0.44''$).

Because the observed flux distribution can be reasonably well fit by a model in which the intrinsic flux distribution is circularly symmetric in the equatorial plane, we have not studied any models in which the intrinsic distribution is not circularly symmetric. Such models are not implausible though, and could probably fit the data equally well. If photoionization is the ionization mechanism in the nucleus of NGC 7052, shielding of the ionization source by the slightly asymmetric dust distribution would lead to a slightly asymmetric flux distribution. The flux would be highest on the side with the smallest dust optical depth, in agreement with the observations.

**Figure 10.** The Hα emission-line flux observed in our major and minor axis spectra, as function of galactocentric distance. The vertical scale is in arbitrary units. The curves are the best (seeing convolved and spatially binned) fits of a circularly symmetric flux distribution in the equatorial plane, with $F(R)$ as in equation (4) and $i = 70°$. The dashed curve assumes there is no absorption of line emission by dust. The solid curve does take into account the absorption of line emission by dust ($F_{\rm obs} = Fe^{-\tau}$, with $\tau$ as in the right panel of Figure 8). The latter model reasonably well explains the asymmetries in the observed flux distribution as being due to dust absorption.

## 6 DYNAMICAL MODELS

### 6.1 Modelling the stellar kinematics

The stellar velocity dispersion determined in Section 4.1 constrains the mass-to-light ratio of the stellar population.

We assume the stellar phase space distribution function to be of the particular form $f(E, L_z)$, where $E$ is the energy and $L_z$ the angular momentum around the symmetry axis. The predicted stellar kinematics can then be calculated from the Jeans equations. We did this using the technique and software described in van der Marel et al. (1994b), taking the seeing, slit width and spatial binning of the data into account. Our standard model requires a mass-to-light ratio $(M/L)_R = 5.0 \pm 1.0$ (in solar units) to reproduce the observed velocity dispersion of $266 \pm 26$ km s$^{-1}$ between $-5''$ and $5''$. This value falls on the relation between mass-to-light ratio and luminosity obtained by van der Marel (1991) for a sample of 37 bright elliptical galaxies. The inferred mass-to-light ratio depends only little on either $\alpha$ or the possible presence of a central black hole, since these influence only the dynamics of the stars very close to the nucleus (e.g., see Figure 11 below).

Our data do not constrain the stellar dynamical structure of NGC 7052, because we have not measured the variation of the stellar kinematics as function of position on the sky. Our mass-to-light ratio estimate can be slightly in error if the dynamical structure differs from that of an $f(E, L_z)$ model. In general, three-integral or triaxial models are more appropriate. However, $f(E, L_z)$ models are much easier to construct, often provide reasonable (though not perfect) approximations to the dynamics of real galaxies (van der Marel 1991), and at least take proper account of flattening.

### 6.2 Modelling the gas kinematics

The circular velocity in the equatorial plane, $V_c(R)$, satisfies

$$V_c^2(R) = R\frac{d\Phi}{dR}, \qquad (11)$$

where $\Phi$ is the gravitational potential of the system. We assume that the potential consists of contributions from the stellar mass density ($\Phi_*$) and a possible central black hole: $\Phi = \Phi_* - GM_{\rm BH}/r$. The total mass of the ionized gas in elliptical galaxies is typically of the order of $10^3 - 10^4$ M$_\odot$ (Phillips et al. 1986), so we neglect its contribution to the potential. The stellar mass density in our models, $\rho = (M/L)j$, is spheroidal with constant flattening. The components of the force $\vec{\nabla}\Phi_*$ exerted by such a mass density can be expressed as one-dimensional quadratures (Binney & Tremaine 1987, equation [2-88]) that are easily evaluated numerically. There is thus no need to make the simplifying assumption that the potential is spherical, as is often done (e.g., Caldwell, Kirshner & Richstone 1986).

In an axisymmetric potential cold gas will eventually settle in the equatorial plane (e.g., Tohline, Simonson & Caldwell 1982). In triaxial potentials the case is more complicated (e.g., de Zeeuw & Franx 1989). Near the nucleus the settling time is generally much shorter than the Hubble time, although this need not be the case in the outer parts (see de Zeeuw 1994 for a review). Here we assume that NGC 7052 is axisymmetric and the gas has indeed settled and reached equilibrium. Hence, it will move on circular orbits with velocity $V_c(R)$. Figure 11 shows the line-of-sight velocity component $v_{\rm los} = V_c(R)\sin i$, as function of projected distance along the major axis, for models with several different black hole masses. If present, the black hole dominates the potential close to the centre and induces a Keplerian rise in the circular velocity, $V_c(R) \propto R^{-1/2}$ for $R \to 0$.

In addition to its gravitational motion, the gas has a (one-dimensional) thermal velocity dispersion

$$\sigma_{\rm thermal} = \sqrt{\frac{kT}{m}}, \qquad (12)$$

**Figure 11.** The mean line-of-sight velocity of the gas along the major axis, $v_{\rm los} = V_c(R)\sin i$ (in km s$^{-1}$), for our standard model. The $(M/L)$ of the stellar population is fixed as determined in Section 6.1. Predictions are shown for several different masses of a possible nuclear black hole : $M_{\rm BH} = 0$, $1.5 \times 10^8$ M$_\odot$, $5 \times 10^8$ M$_\odot$ and $1.5 \times 10^9$ M$_\odot$. The abscissa is in pc, with $1''$ on the sky indicated by an arrow.

where $k$ is Boltzmann's constant, and $m$ is the average mass of a gas particle. The typical temperature of the ionized gas-component in LINERS is $\sim 10^4$ K (Osterbrock 1989), and $\sigma_{\rm thermal}$ is thus on the order of $\sim 10$ km s$^{-1}$.

The gas might also have a non-thermal velocity dispersion. We will loosely refer to such a contribution as 'turbulent', although we will not attempt a physical description of the motions. We merely assume any turbulence to be isotropic, and parametrize the (one-dimensional) turbulent velocity dispersion phenomenologically as

$$\sigma_{\rm turb}(R) = \sigma_0 \times \exp(-R/R_t), \qquad (13)$$

where $R_t$ is the scale length of the turbulence. Turbulent motions could arise through, e.g., shocks or interaction of the gas with the radio jet of NGC 7052. Note in this context that, although LINERS are generally thought to be ionized by a non-thermal continuum, shocks have often been discussed as alternative or additional source of ionization. See Ho, Filippenko & Sargent (1993) and Maoz et al. (1995) for recent reviews and discussions on this subject.

The thermal and turbulent velocity dispersion need not disturb the bulk flow of the gas on circular orbits. We therefore assume the local velocity distribution of the gas, $\mathcal{L}(R,v)$, to be Gaussian with mean line-of-sight velocity $v_{\rm los} = V_c(R)\sin i$, and line-of-sight velocity dispersion $\sigma_{\rm los}(R) = \sqrt{\sigma_{\rm thermal}^2 + \sigma_{\rm turb}^2(R)}$:

$$\mathcal{L}(R,v) = \frac{1}{\sqrt{2\pi\sigma_{\rm los}^2(R)}} \exp\left(\frac{-[v - V_c(R)\sin i]^2}{2\sigma_{\rm los}^2(R)}\right). \qquad (14)$$

The gas could also have a non-zero gravitational (bulk) dispersion. This would be the case if, e.g., the gas resided in individual clouds orbiting on non-circular orbits in the gravitational potential of the system (as suggested by, e.g., Cinzano & van der Marel 1994, in their models for NGC 2974). Although physically quite different from turbulence, it leads to similar predictions for the observed gas kinematics. The only difference is that the predicted mean streaming velocities are somewhat lower, due to the effect of 'asymmetric drift' (e.g., Binney & Tremaine 1987). We show below that models with turbulent velocity dispersions fit the observed rotation curve quite well. In the following, models with non-zero bulk velocity dispersion for the gas are therefore not discussed.

For any particular position along the slit we adopt a Monte-Carlo procedure to take into account the effects of seeing, the slit width and pixel size. At each Monte-Carlo step a random position is drawn from the rectangular region on the sky defined by the slit width and pixel size, and a two-dimensional vector is added, drawn at random from the PSF. The line-of-sight velocity distribution at this position (eq. [14]) is then calculated, and added to the sum of the previous Monte-Carlo steps, weighted with the flux (taking dust absorption into account). Some $10^5$ steps suffice to obtain accurate results. The result is convolved with a Gaussian of the appropriate width to model the instrumental velocity resolution. This then yields the predicted emission line profile shape as function of velocity.

As a brief aside, let us stress here the importance of modelling the seeing. Consider for example the case of a galaxy with a homogeneous mass density core, and with constant $M/L$. The slope of the circular velocity curve is then proportional to the core density. Seeing flattens the slope of the observed rotation curve near the nucleus. So if the core density is determined from the observed slope without modelling the seeing, it is underestimated. As a result, it will be inferred incorrectly that $M/L$ has a central minimum. This effect might be responsible for the increase in $M/L$ with radius reported by Caldwell et al. (1986) for NGC 7097. This result was previously questioned by Bertola et al. (1991), who suggested it to be due to gas motions on elliptic orbits in a triaxial potential.

### 6.3 Gas rotation velocities and velocity dispersions

Figure 12 shows the rotation velocities and velocity dispersions predicted by our model, for H$\alpha$ and [NII] separately, along both the major and minor axes. Solid curves are for the case when there is no central black hole and no gas turbulence. The major axis rotation curves are extremely well fit. The predicted slope of the rotation curve depends on the stellar mass near the nucleus, which is determined mainly by $\alpha$. The value $\alpha = -1.3$ yields the best fit, which is why we adopted this value for our standard model. The fit is less good for other values of $\alpha$, although one could improve the fits of such models by choosing a slightly different value for the stellar mass-to-light ratio (assuming that the estimate in Section 6.1 would not be correct). Either way, no black hole is required to fit the steep major axis rotation curve. However, the model fails to explain the central peak in the observed gas velocity dispersions. This was found to be the case independent of the particular choice for $\alpha$ or the mass-to-light ratio. To fit the observed velocity dispersions of the gas one needs to invoke the presence of a massive black hole, gas turbulence, or a combination of both.

Seeing convolution of a Keplerian rotation curve around a massive black hole increases the observed central velocity dispersion. Dashed curves in Figure 12 show the velocity dispersions for NGC 7052 predicted by models with different black hole masses. The observed dispersions are reasonably well fit when $M_{\rm BH} = 10^9 \, M_\odot$. Inclusion of a black hole does not alter the predicted rotation velocities significantly, because the Keplerian part of the circular velocity curve is much smaller than the seeing radius (cf. Figure 11).

**Figure 12.** Rotation velocities and velocity dispersions for H$\alpha$ and [NII], along the major and minor axes. The solid curves are the predictions of our standard model, without gas turbulence and without a central black hole. The major axis rotation curves are well fit. The predicted minor axis rotation velocities are slightly non-zero because of the dust absorption in the model. The model does not fit the observed central peak in the major axis velocity dispersion. This can be remedied by invoking gas turbulence or the presence of a central black hole. The dashed curves in the bottom panels show the effect of adding a black hole of mass $1.5 \times 10^8 \, M_\odot$, $5 \times 10^8 \, M_\odot$ and $1.5 \times 10^9 \, M_\odot$, respectively. The central velocity dispersion for [NII] is higher than for H$\alpha$, because the flux distribution of the former is more centrally concentrated. Inclusion of a black hole in the model does not alter the predicted rotation velocities significantly.

Alternatively, one can fit the observed velocity dispersions by invoking gas turbulence of the form (13). The observed width of the central peak in the gas velocity dispersions sets an upper limit on the scale length of the turbulence $R_t$. Good fits to the data can be obtained for all values of $R_t$ smaller than the seeing FWHM of the major axis observations. If $R_t = 0.5''$, one needs a central turbulence $\sigma_0 \approx 350 \, \rm km \, s^{-1}$. If $R_t$ is equal to the scale length of the unresolved part of the flux distribution ($0.1''$, see Section 5.5), one needs $\sigma_0 \approx 1000 \, \rm km \, s^{-1}$.

To examine whether turbulent motions on the order of $1000 \, \rm km \, s^{-1}$ are realistic, we calculated the central escape velocity for our standard model without a black hole. A general expression for this quantity for models with luminosity densities of the form (2) is given by Qian et al. (1995). The central escape velocity is $1459 \, \rm km \, s^{-1}$. It is finite in spite of the infinite mass of our model. Turbulent motions exceeding or even close to the central escape velocity are unlikely to exist. This argument favours models with smaller $\sigma_0$ and larger $R_t$.

The models with either a central black hole or turbulence both predict a higher central velocity dispersion for [NII] than for H$\alpha$, as is observed. This is because the flux distribution of [NII] is more peaked towards the nucleus than that of H$\alpha$. Consequently, the observed [NII] flux in the central pixel originates on average from a region nearer to the nucleus (where the RMS velocity is higher) than the observed H$\alpha$ flux.

### 6.4 Emission line profiles

The observed gas velocity dispersions can be fit either by invoking the presence of a central black hole, or by invoking gas turbulence. In an attempt to discriminate between these models we now consider the predicted and the observed emission line profile *shapes*. The observed line profiles away from the centre are slightly asymmetric with a tail at low rotation velocities (Figure 5). This is caused by seeing convolution, and is reasonably well reproduced by all models that fit the major axis rotation curve. Of more interest here is the shape of the emission line profile in the very centre.

Figure 13 compares model predictions (thick curves) and observations (thin curves) for the observed central H$\alpha$ emission line. The left panel shows the predictions for the model with a $10^9 \, M_\odot$ black hole. The predicted emission line shape is completely different from the observed line shape and we hence rule out this model. The predicted line shape has three local maxima. These can be understood in terms of the circular velocity curve of the model (Figure 11). Due to the Keplerian rise at small radii, there is no gas along the major axis with $|v_{\rm los}| < v_{\rm min} \approx 300\,{\rm km\,s}^{-1}$. This results in two local maxima in the predicted line shape at velocities $\sim\pm v_{\rm min}$. The predicted flux at these two local maxima is not the same in this particular case, because: (i) the observed profile is slightly offset from the nucleus ($r = -0.05''$); and (ii) the model takes into account absorption by dust, which is asymmetrically distributed. The third local maximum in the line profile, at $v \approx 0\,{\rm km\,s}^{-1}$, is due to the gas outside the region influenced by the black hole, which is roughly in solid body rotation. If the black hole mass were reduced to zero, this would be the only remaining maximum.

**Figure 13.** The thin curve in each panel is the observed Hα emission line (as in Figure 5) for the galaxy centre ($r = -0.05''$). The thick curves show the predictions of our models. Left: $M_{\rm BH} = 10^9\,M_\odot$, and no gas turbulence; Middle: gas turbulence with $\sigma_0 = 350\,{\rm km\,s}^{-1}$ and $R_t = 0.5''$, and no central black hole; Right: Gas turbulence with $\sigma_0 = 300\,{\rm km\,s}^{-1}$ and $R_t = 0.5''$, and also a $5 \times 10^8\,M_\odot$ central black hole. The vertical scale is in arbitrary units. The latter two models both provide an acceptable fit.

The middle panel of Figure 13 shows the predicted line shape for the model with a turbulent gas dispersion with $\sigma_0 = 350\,{\rm km\,s}^{-1}$ and $R_t = 0.5''$. The shape is smooth and close to Gaussian, and fits the observed shape well. Hence, this model is consistent with the data.

The left panel of Figure 13 indicates that any viable model with a black hole must have gas at velocities smaller than $v_{\rm min}$ (as defined above), to avoid the prediction of local maxima in the line profile shape (which are not observed). This can be achieved by constructing combined models, in which the gas near the nucleus is in Keplerian motion around a black hole, but in addition has turbulence of order $v_{\rm min}$. We have constructed several of these combined models. An adequate fit to the observed velocity dispersions can be obtained with, e.g., $M_{\rm BH} = 5 \times 10^8\,M_\odot$ and turbulence with $\sigma_0 = 300\,{\rm km\,s}^{-1}$ and $R_t = 0.5''$. The right panel of Figure 13 shows the central line profile predicted by this model. It fits the data as well as the model in the middle panel, which has turbulence but no black hole.

### 6.5 Predictions for HST observations

Our ground-based observations show that the motions of the ionized gas in NGC 7052 cannot be purely circular, but must have a significant random (turbulent) component near the centre. A central black hole of mass $M_{\rm BH} \lesssim 5 \times 10^8\,M_\odot$ might or might not be present.

High spatial resolution spectra with the HST could provide more stringent constraints on the presence of a black hole. We calculated predicted rotation velocities and velocity dispersions for hypothetical observations with the $0.26''$ diameter circular aperture of the Faint Object Spectrograph (FOS) on the HST. Figure 14 shows the results for the models discussed in Sections 6.3 and 6.4, as function of position along the major axis. Models with a black hole predict higher rotation velocities than those without a black hole. One does not expect to observe rotation velocities as high as have been observed in M87 (Harms et al. 1994), because M87 is closer by than NGC 7052, and has $M_{\rm BH} \approx 2.4 \times 10^9\,M_\odot$.

## 7 CONCLUSIONS

We have presented high spatial resolution broad-band imaging, narrow-band imaging and spectroscopy for the central region of the E4 radio galaxy NGC 7052. The narrow-band image and spectra show that the (previously unresolved) ionized gas near the nucleus resides in a rapidly rotating disc with radius $\sim 1.5''$. The emission line spectrum is characteristic of that of a LINER. We have determined the kinematic quantities and emission line shapes of the gas along both the major and minor axes. The observed major axis rotation velocity rises to $\sim 250\,{\rm km\,s}^{-1}$ at $r = 1.5''$. The velocity dispersion of the ionized gas is $\sim 210 - 260\,{\rm km\,s}^{-1}$ in the centre and $\sim 60\,{\rm km\,s}^{-1}$ at $r = 1.5''$. The central velocity dispersion for [NII] is higher than for Hα, because the flux distribution of [NII] is more centrally concentrated than that of Hα. Our broad-band image confirms the previously reported presence of a nuclear dust lane, extending from $r \approx -1.6''$ to $1.6''$ along the major axis.

We have constructed detailed models to interpret the data. The galaxy is assumed to be axisymmetric with constant flattening, with the gas and dust in the equatorial plane. The gas distribution is assumed to be circularly symmetric. We allow for the possibility of absorption of the emission-line flux by the dust. The optical depth distribution of the dust on the sky, $\tau(x, y)$, cannot be inferred directly from the observed broad-band image, since it depends on the (unknown) fractions of the observed intensity that originated in front of and behind the dust 'screen', respectively. However, $\tau(x, y)$ is fixed uniquely by the observations once a model for the three-dimensional light distribution has been specified. We present a new algorithm for recovering $\tau(x, y)$, that deconvolves simultaneously for the effects of seeing.

**Figure 14.** Predicted rotation velocities and velocity dispersions along the major axis for observations with the 0.26″ diameter circular aperture of the FOS on the HST. The three models of Figure 13 are shown, as well as the model with no black hole and no turbulence. Models with a black hole discern themselves by predicting larger rotation velocities. HST observations can thus provide more stringent constraints on the presence of a black hole in NGC 7052.

The observations of the ionized gas disc, when corrected for the effects of seeing, indicate an inclination angle $i \approx 70°$. The flux distribution of the ionized gas is extremely centrally concentrated. The flux inside the seeing radius is unresolved, hinting at the presence of a narrow-line region in this galaxy. The central cusp-steepness $\alpha$ of the stellar light distribution ($j \propto r^{-\alpha}$ for $r \to 0$) is not well constrained by our data, because of the absorption by dust near the centre. Motivated by a variety of arguments we choose $\alpha = -1.3$ in our 'standard model'. The observed dust morphology is then consistent with a roughly circularly symmetric dust distribution in the equatorial plane, with maximum optical depth in the centre; i.e., a disc, rather than a ring or filament.

To model the gas kinematics, we assume the gas to move on circular orbits in the equatorial plane. The local velocity dispersion of the gas is assumed to be the sum of a thermal and a turbulent (or otherwise non-gravitational) component. The circular velocity is calculated from the combined gravitational potential of the stars and a possible nuclear black hole. The average mass-to-light ratio of the stellar population is assumed to be constant. Its value is determined by fitting to the stellar velocity dispersion determined from our data, assuming a stellar phase-space distribution function of the form $f(E, L_z)$.

The model with no black hole and no turbulence fits the major axis gas rotation curve, but fails to fit the observed central peak in the gas velocity dispersion. This indicates that there must be a strong increase in the RMS gas motion inside the seeing radius. A model with a $10^9 \, M_\odot$ black hole and no turbulence yields a good fit to the observed velocity dispersions, due to seeing convolution of the Keplerian part of the rotation curve. However, this model is ruled out because it predicts a nuclear emission line profile that has two pronounced peaks at $v \approx \pm 300 \, \text{km s}^{-1}$, which are not observed. Models without a black hole, but with strong turbulence ($\gtrsim 300 \, \text{km s}^{-1}$) within the seeing radius, can fit not only the rotation curve and velocity dispersions, but also the observed line profile shapes. Combined models with both a black hole and strong turbulence can also fit the data, but the black hole mass must be $\lesssim 5 \times 10^8 \, M_\odot$. Predictions for future observations with the HST show that such observations should yield more stringent constraints on the mass of a possible black hole. Any black hole with mass $\gtrsim 10^8 \, M_\odot$ will reveal itself by causing the rotation curve observed with HST to be significantly steeper than that reported here.

A complication not addressed here is the possibility of elliptic gas orbits in a triaxial potential. Ideally one would like to include this in the modelling. This has in fact been done successfully for extended ionized gas disks in several early-type galaxies (Bertola et al. 1991; Amico et al. 1993). However, information on the two-dimensional velocity field on the sky is required to place meaningful constraints on such models. Such information is not available for the galaxy we study here, and will generally be very difficult to obtain for any gas disk restricted only to the nuclear region.

Although the upper limit we have set here on the mass of a black hole in NGC 7052 is not very stringent, it is already ∼5 times smaller than the black hole mass inferred for M87 from HST data (Harms et al. 1994). It will be worthwhile to search for and observe other galaxies with nuclear gas discs that are closer by than NGC 7052. Such galaxies can be studied at higher spatial resolution, as measured in physical units, allowing the detection of less massive black holes. One galaxy for which HST observations are planned is NGC 4261. HST photometry of this galaxy in the Virgo cluster have revealed a beautiful nuclear dust disc (Jaffe et al. 1993). Emission line spectra obtained with the WHT show a broad emission-line component, which hints at the presence of a $\sim 4 \times 10^7$ $M_\odot$ black hole. The observed nuclear emission line shapes are not double (or triple) peaked, so some form of gas turbulence is again required (Jaffe et al. 1994). In the near future, kinematic observations of nuclear gas discs are likely to become a widely used tool in the search for massive black holes in galactic nuclei. The modelling and analysis techniques developed here will be useful for interpreting such data.

**ACKNOWLEDGMENTS**

The results in this paper are based on observations obtained with the William Herschel Telescope operated on the island of La Palma by the Royal Greenwich Observatory in the Spanish Observatorio del Roque de los Muchachos of the Instituto de Astrofisica de Canarias, and on observations kindly obtained for us at the Danish Telescope of ESO at La Silla by Marcella Carollo. We studied the literature on NGC 7052 using the NASA/IPAC Extragalactic Database (NED), operated by JPL, Caltech, under contract with NASA. We thank Frank Israel, Eric Emsellem and Tim de Zeeuw for valuable discussions, and Dave Carter for assisting with the WHT observations. FCvdB acknowledges support, #782-373-055, from the Netherlands Foundation for Astronomical Research (ASTRON) with financial aid from the Netherlands Organization for Scientific Research (NWO). In the later stages of this research, RPvdM was supported by NASA through a Hubble Fellowship, #HF-1065.01-94A, awarded by the Space Telescope Science Institute which is operated by AURA, Inc., for NASA under contract NAS5-26555.